\documentclass[amsmath,amssymb, superscriptaddress, preprint]{revtex4-1}

\usepackage[T1]{fontenc}
\usepackage[ansinew]{inputenc}
\usepackage[english]{babel}
\usepackage{amsfonts}
\usepackage{amsmath}
\usepackage{array}
\usepackage{amsthm}
\usepackage{amssymb}
\usepackage{graphicx}
\usepackage{subfigure}
\usepackage{braket}
\usepackage{eucal}
\usepackage{verbatim}
\usepackage[table]{xcolor}
\usepackage{caption}

\usepackage[pdftex,plainpages=false,colorlinks,hyperindex,bookmarksopen,linkcolor=black,citecolor=blue,urlcolor=blue]{hyperref}

\usepackage{braket}

\newcommand{\be}{\begin{eqnarray}}
\newcommand{\ee}{\end{eqnarray}}
\renewcommand{\d}{\mbox{${\rm d}$}} 

\newcommand{\R}{\mathbb{R}}  
\def\eg{{\it e.g. }} 
\def\ie{{\it i.e. }}

\newcommand{\wh}[1]{\widehat{#1}}

\newcommand{\lp}{\ell_{\rm P}}
\newcommand{\mpl}{m_{\rm P}}

\newcommand{\Rh}{R_{\rm H}}
\newcommand{\hRh}{\wh{R} _{\rm H}}
\newcommand{\expec}[1]{\mbox{$\langle\, #1\,\rangle$}}

\begin{document}

\title{Horizon quantum fuzziness for non-singular black holes}

	\author{Andrea Giugno}
		\email{A.Giugno@physik.uni-muenchen.de}
		\affiliation{Arnold Sommerfeld Center, Ludwig-Maximilians-Universit\"at, Theresienstra\ss e~37,~80333 M\"unchen, Germany.}
	\author{Andrea Giusti}
		\email{agiusti@bo.infn.it}
		\affiliation{Arnold Sommerfeld Center, Ludwig-Maximilians-Universit\"at, Theresienstra\ss e~37,~80333 M\"unchen, Germany.}
		\affiliation{Dipartimento di Fisica e Astronomia, Universit\`a di Bologna, Via Irnerio~46, 40126 Bologna, Italy.}	
 		\affiliation{I.N.F.N., Sezione di Bologna, IS - FLAG, via B.~Pichat~6/2, 40127 Bologna, Italy.}
	\author{Alexis Helou}
		\email{alexis.helou@physik.uni-muenchen.de}
 		\affiliation{Arnold Sommerfeld Center, Ludwig-Maximilians-Universit\"at, Theresienstra\ss e~37,~80333 M\"unchen, Germany.}

	\date  {\today}

\begin{abstract}
We study the extent of quantum gravitational effects in the internal region of non-singular, 
Hayward-like solutions of Einstein's field equations according to the formalism 
known as Horizon Quantum Mechanics.
We grant a microscopic description to the horizon by considering a huge number
of soft, off-shell gravitons, which superimpose in the same quantum state, 
as suggested by Dvali and Gomez. In addition to that, the constituents of such a configuration
are understood as loosely confined in a binding harmonic potential.
A simple analysis shows that the resolution of a central singularity through quantum physics
does not tarnish the classical description, which is bestowed upon 
this extended self-gravitating system by General Relativity.
Finally, we estimate the appearance of an internal horizon as being negligible, because of the suppression
of the related probability caused by the large number of virtual gravitons.
\end{abstract}


\maketitle

\section{Introduction}
The theory of General Relativity (GR) still presents some major physical problems,
despite having maintained a high degree of consistency with astrophysical observations
for more than a century. 
One of such issues arises in a merely classical framework. In fact, the existence of a trapped
surface implies geodesic incompleteness in a globally hyperbolic
system, which is solution of Einstein's field equations sourced by some reasonable
matter distribution (\eg one that obeys the null energy condition).
This is the content of the so-called \emph{Singularity theorems} in a nutshell: nothing prevents
the formation of a curvature singularity at some stage of the gravitational collapse~\cite{Oppenheimer}, 
at least in the spherically symmetric 
case \cite{ft1}.
One may logically argue that this setback stems from the failure of classical physics
at describing short-scale phenomena, for which a fully quantum apparatus is required.
Unfortunately, any attempt at quantising GR bumps inevitably into various inconsistencies,
which lead to the infamous UV-incompleteness of the theory
(for more information, see \eg~\cite{Malafarina:2017csn} and references therein).
\par
A sensible argument to tackle down this premise is to assume that our way of modeling gravity
should be extensively modified, as we get closer and closer to regions of extreme curvature.
In view of this statement, GR should be regarded as a low-energy effective theory of gravity, 
which loses its reliability beyond a certain energy scale.
Therefore, it is of paramount importance to study the extent of such corrections,
for they can provide some hints at a bottom-up construction 
of a self-consistent quantum theory of gravity. 
\par	
One of the main ideas emerging from almost every tentative model of Quantum Gravity (QG) is the appearance of a \textit{critical} energy scale $\Lambda$ (and a corresponding length scale~\cite{ft2}  
$\ell \sim \hbar / \Lambda$) above which the quantum nature of gravity cannot be completely neglected any longer. 
In other words, one expects quantum gravitational effects to modify the geometric description
of the spacetime, when the scalar curvature $\mathcal{R} \sim \ell ^{-2}$. 
This feature should be accordingly matched by suitable changes in the structure 
of the metric tensor, as long as the classical metric description is preserved.
However, it is important to remark that $\ell$ does not coincide with the QG regime $\ell _{QG}$, 
but rather it is just the scale at which some corrections to GR
come into play in the form of appropriate effective metric tensors. 
Here we still assume that $\ell _{QG} \ll \ell$, in order to limit ourselves to simple modifications
of the field equations, while avoiding to deal with full-fledged quantum 
degrees of freedom of gravity in the process.
\par
The first proposal for a non-singular solution of the field equations of GR equipped with an event horizon dates back to the Sixties, with the seminal paper \cite{Bardeen} by Bardeen. 
As stressed in \cite{Ansoldi}, the Bardeen black hole is an example 
of a more general class of solutions of the gravitational field equations in which the regular interior is obtained by assuming a vacuum-like equation of state for the matter content below a certain length scale, 
in strict analogy with the previous discussion. 
This approach, when applied to static spherically symmetric cases, leads 
to regular black hole solutions with a de Sitter-like core and makes it reasonable to assume 
that a proper UV-complete theory of gravity should lead 
to some physically fundamental constraints on the values of the curvature invariants~\cite{Frolov:1988vj}. 
It is therefore important to stress that the metric studied here (and proposed by Hayward \cite{Hayward})
actually satisfies the said limitations (see \eg~\cite{Frolov-NSBH}).
\par
 The purpose of this paper is to analyse some semiclassical aspects of the Hayward spacetime by means of the horizon quantum mechanics formalism~\cite{Casadio:2013tma,Casadio:2015qaq,Casadio:2016fev}, while working within the framework of the corpuscular model for quantum black holes~\cite{D&G1,D&G2}.
\par
This paper is therefore organised as follows:

	In Section \ref{sec-1} and \ref{sec-2}, we first review some general aspects of the geometry taken into consideration and we introduce all the main operational features of the horizon quantum mechanics.
	
	Consequently, in Section \ref{sec-3}, we first review the corpuscular model for black holes, together with its connection with harmonic black hole model, and we then compute the probability for $N$ soft, virtual gravitons to be confined either on the internal or the external horizon.
	
	Finally, we conclude the paper with remarks and hints for future research. 

\section{Geometry of Hayward's solution} \label{sec-1}
A pivotal example of static non-singular black hole, belonging to the class of solutions 
of the Einstein's equations discussed above, is the so-called \textit{Hayward spacetime}, 
defined by the line element
\be 
ds^2 = - f(r) \, dv ^2 + 2 \, dv \, dr + r^2 \, d \Omega ^2 \ ,
\ee
with
\be 
f(r)
= 
1 - \frac{2 \lp M \, r^2/\mpl}{r^3 + 2\lp  M \, \ell ^2/\mpl} 
\ .
\label{eq:MetrFuncAdv}
\ee
In the above, $v$ is the \textit{advanced time} coordinate, $r$ the areal radius, 
$d \Omega ^2 = d \theta ^2 + \sin ^2 \theta \, d \varphi $ the volume element of a $2$-sphere, 
$M$ the ADM mass and $\ell$ represents the critical parameter arising from the corresponding UV-complete theory of gravity. 
In particular, this model is built in such a way that we recover the asymptotic fall off 
of the Schwarzschild metric as $r \to \infty$ and a ``de Sitter-like'' behavior 
\be 
f(r) 
\sim 
1 - \frac{r^2}{\ell ^2} 
\ , 
\quad 
\mbox{as} \,\, 
r 
\to 
0 
\ .
\ee
In addition to that, the corresponding Einstein's field equations reduce to a simple form,
$G_{\mu \nu} \sim - (3/\ell ^2) \, g_{\mu \nu} $, near the origin. 
For further details on the topic, we invite the interested reader to refer to~\cite{Frolov-NSBH, Hayward}.
\par
The trapping horizons of the Hayward spacetime are given by the zeros of the equation $\nabla_{\mu}r\nabla^{\mu}r=0$, that is by the ones of the function $f(r)$
in Eq.~\eqref{eq:MetrFuncAdv}. 
This equation reads
\be
 r^3 -2\frac{\lp M }{\mpl}\, r^2+2\frac{\lp M }{\mpl}\, \ell^2=0 \, .
 \label{eq:fr=0}
\ee 
In particular, in the following we will be interested in those configurations that allow for the formation of two distinct horizons (such as in the Kerr or Reissner-Nordstr\"om cases for example).

	Now, the left hand side of Eq.~\eqref{eq:fr=0} is a cubic polynomial. The fundamental theorem of algebra states that it will admit three (generally) complex roots, which we can denote by $\Rh^>$, $\Rh^-$ and $\Rh^<$. Then, from Vieta's theorem we get the following relations
	\be
 && \Rh^> + \Rh^- + \Rh^< 
 \, = \, 
 \frac{2\lp M}{\mpl}
 \, > \,
 0 
 \label{eq:Vieta1} 
 \ ,
 \\
 && \Rh^> \, \Rh^- + \Rh^> \, \Rh^< + \Rh^- \, \Rh^< 
 \, = \,
 0 
 \label{eq:Vieta2}
 \ ,
 \\
 && \Rh^> \, \Rh^- \, \Rh^< 
 \, = \,
 2\frac{\lp M }{\mpl}\, \ell^2 
 \,< \,
 0 
 \ .
 \label{eq:Vieta3}
\ee 
If $\Rh^> , \, \Rh^- , \, \Rh^< \in \R$, it is easy to infer from Eq.~\eqref{eq:Vieta3} 
that either one or all of the roots are negative, 
whereas Eq.~\eqref{eq:Vieta1} forbids the latter possibility. Therefore, if the three roots are real, one must have two positive roots and a negative one. If this is the case, we then denote the negative solution with $\Rh^-$ and the other two with $\Rh^> , \,  \Rh^<$ according to the condition $\Rh^> \geq  \Rh^<$.
\par
If we perform the change of variable $z=r - (2 \lp M / 3\mpl)$ in \eqref{eq:fr=0} we get
\be
 z^3 -\frac{4\lp^2M^2}{3\mpl^2}z
  -\frac{16\lp^3 M^3}{27\mpl^3} + \frac{2\lp M}{\mpl}\, \ell^2
 \, = \,
 0 
\ ,
\ee
to which we can then apply the Cardan method for solving cubic equations.

The discriminant is 
\be
\Delta
\, = \,
\left(\frac{2 \, \lp \,  M}{\mpl}\, \ell\right)^2
\, \left( \frac{16 \ \lp^2 \ M^2}{\mpl^2} - 27 \ \ell^2 \right)
\ ,
\ee 
so that the equation admits at least two distinct real solutions when it is strictly positive,
and two degenerate real solutions if it vanishes. 
Since the coefficients of the polynomial are functions of the ADM mass $M$ and the cut-off $\ell$,
the horizon structure is regulated by a simple inequality, namely
\be
0
\, \leq \,
\frac{3\sqrt{3}\,\mpl}{4 \ \lp}\,\ell
\, \leq \,
M
\label{eq:Bound2Hor}
\ .
\ee 
It is worth noting that the saturation of the r.h.s.~inequality represents the extremal case,
whereas two distinct horizons appear otherwise.

Summarizing, the three solutions are
\be
\label{Rhj}
\Rh ^{(j)}
\,=\,
 \frac{4\lp M}{3\mpl} \cos\left[\frac{1}{3}\arccos \left(1 - \frac{27\mpl^2 \ell^2}{8\lp^2M^2} \right)+\frac{2\pi j}{3}\right] 
 + \frac{2\lp M}{3\mpl} 
 \ ,
\qquad
j
\,=\,
1 ,\, 2 ,\, 3
\ .
\ee
Eq.~\eqref{eq:Bound2Hor} tells us that $\Rh ^{(1)}$ ranges from 
$-\frac{2\lp M}{3\mpl}$ to $0^-$, allowing the identification with $\Rh^-$.
The same procedure tells us that $j=2$ represents $\Rh^<$, since its lower bound is $0$,
whence $j=3$ gives $\Rh^>$, that can grow up to $2 \lp M/\mpl$. Besides, the extremal case corresponds to a value of the radius given by
\be
R_{\rm H,\,ext}
=
\frac{4\lp M}{3\mpl}
\ ,
\ee
which is obtained by setting $\Rh^<=\Rh^>$.
\par
When the internal cut-off $\ell$ is such that $$\ell \ll \frac{\lp M}{\mpl} \, ,$$ 
the metric function~\eqref{eq:MetrFuncAdv} approximatively describes 
a Schwarzschild solution with a small, but (potentially) still macroscopic, inner de Sitter core. In fact, if we recall that
\be 
\arccos \left( 1 - \frac{27}{8} \, x^2 \right) 
\,=\, 
\frac{3 \sqrt{3}}{2} \, x + \mathcal{O} \left( x^3 \right) 
\ , 
\quad \cos x 
\,=\, 
1 - \frac{x^2}{2} + \mathcal{O} (x^4)
\ ,
\ee
for $x \ll 1$, then expanding Eq.~\eqref{Rhj} for $x = \mpl \ell / \lp M \ll 1$ it is easy to see that
\be
\Rh^> \sim \frac{2\lp}{\mpl} \, M - \frac{\mpl \, \ell^2}{2 \, \lp \, M} = R _{\rm s} - \frac{\ell^2}{R _{\rm s}}
\label{Rh>}
\ , 
\ee
\be
\Rh^<  
\sim
\ell + \frac{\mpl \, \ell^2}{4 \, \lp \, M} = \ell + \frac{\ell^2}{2 \, R _{\rm s}}
\ ,
\label{Rh<}
\ee
where $R _{\rm s} \equiv 2 \lp M / \mpl$ is the Schwarzschild radius.
%
	
\section{Horizon Quantum Mechanics} \label{sec-2}
\label{sec:HQM}
In this Section, we shall introduce the reader to some general aspects of the Horizon Quantum Mechanics (HQM), 
according to its most recent updates~\cite{Casadio:2013aua, Casadio:2014twa, Casadio:2014pia, Casadio:2015rwa, Casadio:2017nfg, Giusti:2017byx}.

If one envisages the source of the gravitational field as a purely quantum object
under this perspective, then one can faithfully picture both the ADM mass and the gravitational
radius as quantum variables in return.

It is therefore reasonable to describe the entire system in terms 
of an Hilbert space $\mathcal{H}$ defined as
	\begin{equation}
	\mathcal{H} = \mathcal{H} _{\rm S} \otimes \mathcal{H} _{\rm G}
	\end{equation}
	where $\mathcal{H} _{\rm S}$ and $\mathcal{H} _{\rm G}$ represent the Hilbert spaces for the source and for the geometry, respectively.
	
The quantum state of the source should be specified in terms of an Hamiltonian operator 
$\wh{H} := \wh{H} \otimes \wh{\mathbb{I}} _{\rm G}$, 
with $\wh{\mathbb{I}} _{\rm G}$ the identity on $\mathcal{H} _{\rm G}$, such that
\be 
\wh{H} 
= 
\left( \sum _{\alpha} E_\alpha \, \ket{E_\alpha} \bra{E_\alpha} \right) \otimes \wh{\mathbb{I}} _{\rm G} \, ,
\label{eq:spectrum}
\ee
where $\ket{E_\alpha}$ represent the ``energy'' eigenstates of the source corresponding to the eigenvalues $E_\alpha$ and $\alpha$ is a set of quantum numbers  parametrising its spectral decomposition. 

	In this framework~\cite{Casadio:2016fev, Casadio:2017nfg, Giusti:2017byx}, we depict the wave-function of the whole system $\ket{\Psi} \in \mathcal{H}$ as an entangled state among $\ket{E_\alpha}$ and the eigenstates of the gravitational radius operators acting on $\mathcal{H} _{\rm G}$, \ie
\be
\ket{\Psi}
=
\sum_{\alpha,\beta,\gamma}
C\left(E_\alpha ,\,\Rh{}^>_\beta , \, \Rh{}^<_\gamma \right)
\ket{E_\alpha}\ket{\beta}_>\ket{\gamma}_<
\ ,
\ee
where
\be
\hRh^{(j)} \ket{\beta}_{(j)}
\,=\,
\Rh{}^{(j)}_\beta \ket{\beta}_{(j)}
\ ,
\ee
with the subscript ``$j$'' standing for either ``$>$'' or ``$<$''.

	If we now connect the quantum uplifting of the ADM mass with the spectral decomposition~\eqref{eq:spectrum}, it is easy to see that not all the vectors belonging to $\mathcal{H}$ are physical states. In fact, they are the ones that allow us to gain a \textit{horizon-law relation}, intended as a constraint ``\emph{\`a la Gupta-Bleuler}'', between the spectrum of $\widehat{H}$ and the one of $\hRh^{(j)}$, with $j \in \{>, \, <\}$, namely
\be
\left(
\hRh^> - \wh{\mathcal{O}}^>
\right) \, \ket{\Psi} _{\texttt{phys}}
&=&
0
\label{HorConstr>}
\ ,
\\
\left(
\hRh^< - \wh{\mathcal{O}}^<
\right) \, \ket{\Psi} _{\texttt{phys}}
&=& 
0
\ .
\label{HorConstr<}
\ee
In this case, the operators $\wh{\mathcal{O}}^<$ and $\wh{\mathcal{O}}^>$ are chosen in order to reproduce the approximated relations in Eq.~\eqref{Rh>} and Eq.~\eqref{Rh<}, \ie
\be
\wh{\mathcal{O}}^>
&=&
\frac{2\lp}{\mpl}\,\wh H-\frac{\mpl \ell^2}{2\lp} \, \wh H^{-1}
\label{O>}
\ ,
\\
\hat{\mathcal{O}}^<
&=&
\ell+\frac{\mpl \ell^2}{4\lp} \, \wh H^{-1}
\label{O<}
\ .
\ee
Notice that $\ell$ is regarded as an external parameter. This is reasonable because it should represent some sort of UV cut-off arising from a fundamental theory of quantum gravity rather than from a semiclassical description of the interplay between the quantum source and the geometry of the system.

	Now, if we trace away from $\ket{\Psi}$ the contribution of the geometry, we are left with the wave-function of the collapsed matter, \ie
	\be \label{psis}
	\ket{\Psi} _{\rm S} = \sum _{\alpha} C _{\rm S} (E_\alpha) \, \ket{E_\alpha} \, , \qquad 
	C _{\rm S} (E_\alpha) \equiv C\left(E_\alpha ,\,\Rh{}^> (E_\alpha) , \, \Rh{}^< (E_\alpha) \right) \, ,
	\ee 
	where the relations $\Rh{}^> (E_\alpha)$ and $\Rh{}^< (E_\alpha)$ are provided by the constraints \eqref{HorConstr>} and \eqref{HorConstr<}. This quantum state is then related to the classical ADM mass through the expectation value
\be
M
\, \mapsto \,
\braket{\Psi _{\rm S} | \wh H | \Psi _{\rm S}}
=
\sum_\alpha |C_{\rm S}(E_\alpha)|^2 \, E_\alpha
\ .
\ee

By tracing out the contribution of the source, together with one of the two contributions of the horizons, we can then define two (non-normalized) \emph{horizon wave-functions}, namely  
\be
\Psi_> \left(\Rh^> \right)
\simeq
C\left(E_\alpha (\Rh^>) ,\,\Rh{}^> , \, \Rh{}^< (\Rh^>) \right)
\ ,
\label{hwf>}
\ee
\be
\Psi_< \left(\Rh^< \right)
\simeq
C\left(E_\alpha (\Rh^<) ,\,\Rh{}^> (\Rh^<) , \, \Rh{}^< \right)
\ .
\label{hwf<}
\ee
	
	Then, from these two wave-functions we can build two probability density functions for a lump of matter $\ket{\Psi} _{\rm S}$ to be equipped with an internal horizon located in $\Rh^<$ and an external one in $\Rh^>$, once gravitational collapse has taken place. To be more precise, we have
	\be 
	\mathcal{P} _{>} (r^{>} _{\rm H}) := 4 \, \pi \, {r^{>} _{\rm H}}^2  \, |\Psi _{>} (r^{>} _{\rm H})| ^2 \, ,
	\ee
	\be 
	\mathcal{P} _{<} (r^{<} _{\rm H}) := 4 \, \pi \, {r^{<} _{\rm H}} ^2 \, |\Psi _{<} (r^{<} _{\rm H})| ^2 \, .
	\ee
This allows us to infer whether $\ket{\Psi} _{\rm S}$ sources a black hole. Indeed, denoting the probability density for $\ket{\Psi} _{\rm S}$ to be localized within the external horizon,
\be 
\wp (r < r^{>} _{\rm H}) = P _{\rm S} (r < r ^> _{\rm H}) \, \mathcal{P} _{>} (r^{>} _{\rm H}) \, ,  
\ee
where $P _{\rm S} (r < r ^> _{\rm H}) = 4 \, \pi \int _0 ^{r ^> _{\rm H}} \bar{r} ^2 \, | \Psi _{\rm S} (\bar{r}) |^2 \, 
{\rm d} \bar{r}$, then the probability for $\ket{\Psi} _{\rm S}$ to be a black hole is given by
\be \label{def-pbh}
P_{\rm BH} = \int_0 ^\infty \wp (r < r^{>} _{\rm H}) \, \d r^{>} _{\rm H} \, .
\ee

 In a similar fashion, we can estimate the likelihood for the system to be trapped within the internal horizon. It will be denoted by $P_{\rm IH}$.
\section{Application to non-singular BHs} \label{sec-3}
\label{sec:Corpusc}
%
Before investigating whether the effective inclusion of QG effects in the interior of an event horizon 
preserves or spoils the GR description of gravitating systems, we have to model a 
suitable microscopic theory.
Therefore, we quickly recall the basic features of the corpuscular black hole 
model of Dvali and Gomez~\cite{D&G1, D&G2, Casadio:2016zpl, Casadio:2017cdv, Casadio:2017twg}, which is the setting we are taking into consideration. 
In this way, black holes are extended objects, which hint at a UV completion
of gravity through the production of a huge number $N$ of soft, gapless modes, instead
of a low amount of very hard ones.
\par
By means of a binding potential, these constituents can superimpose 
in one particular quantum state, effectively making the compact object 
a self-gravitating Bose-Einstein Condensate (BEC)~\cite{Ruffini, Colpi, Membrado, Chavanis, Kuhnel1,Kuhnel2}, at least to leading order of approximation.
Moreover, the virtual gravitons are expected to be \emph{marginally bound} in this potential
well, giving rise to the so-called \emph{maximal packing} relation
\be
\mu+U_{\rm N}
\,\simeq\,
0
\ee
expressed in terms of the Newtonian potential energy profile
\be \label{eb}
U_{\rm N}
\simeq
-N \, \alpha \, \mu \,\Theta(\lambda_\mu-r)
\ ,
\ee
where $\mu$ is the graviton effective mass, related to its Compton/de Broglie wavelength by $\lambda _\mu \simeq \mpl \, \lp / \mu$, and $\alpha = \lp ^2 / \lambda _\mu ^2 = \mu ^2 / \mpl ^2$ is the effective coupling constant. 
 
	This set-up is particularly interesting since we can quantize the relevant features of a 
	black hole solely in terms of the mean number of constituents $N$. 
	In fact, as one can easily infer from the energy balance \eqref{eb}, the effective coupling reads $\alpha\simeq 1/N$, while 
\be
M
\,\simeq\,
\sqrt{N} \, \mpl
\ , 
\qquad
\mu
\,\simeq\,
\frac{\mpl}{\sqrt{N}}
\ee
and $\lambda_\mu=\hbar/\mu=\sqrt{N}\,\lp$. In particular, it is also possible to justify 
the latter estimate through a simple argument. 
If a quantum black hole is supposed to establish a bridge 
between quantum mechanics (QM) and GR, the characteristic size of quantum fluctuations, $\lambda_\mu$, 
should match indeed the extent of a typical gravitational radius $R_g\sim\lp \, M/\mpl$.
Moreover, in these terms, the bound~\eqref{eq:Bound2Hor} can be immediately rewritten as
\be
\ell
\,\leq\,
\frac{4}{3\sqrt{3}}\,\frac{M}{\mpl}\,\lp
\,=\,
\frac{4}{3\sqrt{3}}\,\sqrt{N}\,\lp
\ .
\label{eq:Bound2HorCrit}
\ee
\par
Although approximating the binding potential with a square well leads 
to interesting results~\cite{Casadio:2015lis}, 
a more appropriate approximation for such a potential is provided by the harmonic black hole model~\cite{Casadio:2017nfg, Casadio:2013ulk}, for which we have that
$V(r)=V_0(r) \, \Theta(\lambda_\mu-r)$, where
\be
V_0(r)
=
\frac{1}{2} \, \mu \,\omega^2 \, \left(r^2-\lambda_\mu^2 \right) \,
\label{HarmPot}
\ .
\ee
We chose to set the width of the well of order $\lambda_\mu$ in order to enforce the 
connection between the confinement given by the Schwarzschild radius
and quantum physics.
Moreover, the central singularity is resolved by the quantum mechanical
properties of collapsed matter, as it has already been discussed by means of 
post-Newtonian arguments~\cite{Casadio:2016zpl,Casadio:2017cdv}.
\par
Therefore, the relevant quantum states of the toy gravitons in this spherically symmetric
geometry are depicted by the single-particle eigenfunctions
\be
\phi_{nl}(r)
\,=\,
\mathcal{N}_{nl} \,
r^l \,  L_n^{\left(l+1/2\right)}\left(\frac{\omega\,r^2}{\lambda_\mu}\right)
\, \exp \left(-\frac{\omega\, r^2}{2\lambda_\mu}\right)
\label{HarmSol}
\ ,
\ee
where $\mathcal{N}_{nl}$ is a normalisation coefficient and $L_n^a(x)$ are the generalised Laguerre polynomials. These functions are solutions of the Schr\"odinger equation
\be
\frac{\lp^2 \, \mpl^2}{2 \, \mu} \,\left(\triangle-\frac{l(l+1)}{r^2} \right) \, \phi_{nl}(r)
\,=\,
(V_0-E_{nl}) \, \phi_{nl}(r)
\ ,
\ee
under the radial Laplacian in spherical symmetry, \ie
\be
\triangle \phi_{nl}(r)
\,=\,
\frac{1}{r^2}\,\frac{d}{d r}
\,\left[ r^2 \,
\frac{d \phi_{nl}(r)}{d r}
\right]
\ .
\ee
\par
We recall that the energy spectrum reads
\be
E_{nl}
\,=\,
\lp \, \mpl \,\omega \,
\left(
2n +l+\frac{3}{2}
\right) +V_0
\label{Spectrum}
\ ,
\ee
where $V_0=V_0(0)$.
The maximal quantum numbers $n_0$ and $l_0$ designate the most loosely bound state, 
which translates to $E_{n_0,l_0}\approx 0$. 
This concept is particularly useful because, from Eq.~\eqref{HarmPot} and the condition
\be
V_0
\,=\,
-\lp\mpl \, \omega \left(2 \, n_0 +l_0+\frac{3}{2} \right) \, ,
\ee 
one can immediately read off that
\be
\omega
\,=\,
\frac{2}{\lambda_\mu}\,\left(2n_0+l_0+\frac{3}{2}\right) \, .
\label{omega}
\ee

	The energy spectrum Eq.~\eqref{Spectrum} is then
\be
E_{nl}
= 
2\,\mu \, \left(2 \, n_0+l_0+\frac{3}{2}\right) \, \left[2 \, (n-n_0) +(l-l_0)\right]
\ .
\ee
Since the Hayward spacetime describes a static black hole, one can easily show that the (Komar) total angular momentum vanishes. Therefore, we choose to focus on quantum states satisfying $l = l_0 \equiv 0$. 

Although many microscopic configurations may realise this feature, we impose this condition for the sake of simplicity and to avoid the 
unnatural choice of one particular Clebsch-Gordan coefficient among a great number thereof.

Eventually, the quantum state of every toy graviton is decomposed over the 
basis spanned by
\be
\phi_n(r)
 &=& 
\mathcal{N}_n \,  L_n^{\left(1/2\right)}\left(\frac{\omega \, r^2}{\lambda_\mu}\right)
\, \exp \left( -\frac{\omega\,r^2}{2 \, \lambda_\mu} \right) \\
&\simeq&
\mathcal{N}_n \,  L_n^{\left(1/2\right)} \, \left[\frac{(4 \, n_0 + 3)\,r^2}{N \, \lp^2} \right]
\, \exp \left[ - \left(2 \, n_0+\frac{3}{2}\right) \, \frac{r^2}{N \, \lp^2} \right]
\ee
according to the scaling relation for $\lambda_\mu$, Eq.~\eqref{omega} and
\be
\frac{\omega}{\lambda_\mu}
=
\frac{2}{\lambda_\mu^2}\left(2 \, n_0+\frac{3}{2}\right)
=
\frac{4 \, n_0+3}{N\,\lp^2}
\ .
\ee
\subsection{Black hole probability}
The entire system is described, in terms of the single-particle wave-functions $\phi_n^{(i)}$,
by the multi-particle state
\be
\ket{\Phi}
=
\bigotimes ^N _{i=1}  \ket{\phi^{(i)}}
\,=\,
\bigotimes ^N _{i=1} \left[
\sum_{n=0}^\infty c_n \ket{\phi_n^{(i)}}
\right]
\ ,
\label{MultiPart}
\ee
which is an eigenstate of the $N$-particle Hamiltonian according to 
\be
\braket{\Phi \, | \, \wh{H} \, | \, \Phi}
=\
M
\ .
\label{ADM}
\ee
Let us recall that $M=N\mu$ is just the ADM mass.

Clearly, in position space
\be
\braket{r \, | \, \phi_0}
=
\phi_0(r)
\,\simeq\,
\left(
\frac{4 \, n_0 + 3}{\pi \, N \, \lp^2}
\right)^{3/4}
\, \exp\left[ -\left(2 \, n_0+\frac{3}{2}\right) \, \frac{r^2}{N \, \lp^2} \right]
\ .
\ee
Since our purpose is the comparison of the result with a singular case, \emph{i.e.}~Schwarzschild ($\ell=0$), we can afford to model the multi-particle quantum state in the 
simplest possible way.
That is to say, we assume that most of the contribution to the probability
densities comes from the harmonic states~\eqref{HarmSol} carrying
the smallest $n$, \emph{i.e.}~$c_n^{(i)}\sim \delta_{n0}$, $\forall$~$i=1,\dots,N$.
Therefore, we can reliably approximate the multi-particle state~\eqref{MultiPart} 
with the product
\be
\ket{\Phi}
\,\simeq\,
\bigotimes^N_{i=1} \ket{\phi_0^{(i)}}
=
\Bigl[\ket{\phi_0}\Bigr]^N
\ .
\label{state}
\ee

	As discussed in~\cite{Casadio:2017nfg}, it is possible to estimate the probability
for this quantum source to be a black hole by
\be
P_{\rm BH} (N) = \int_0 ^\infty \wp \left( r_1<\Rh \ ,\dots \ , r_N<\Rh \right) \, \d \Rh \, ,
\ee
in a simple way. In fact, from Eq.~\eqref{state} it is easy to see that
\be
P_{\rm S} \, \left( r_1<\Rh\ ,\dots \ , r_N<\Rh \right)
=
\prod_{i=1}^N
P_{\rm S} \left( r_i<\Rh \right)
\ee
and the horizon wave-function for the outer horizon gives rise to a delta-shaped \emph{total} probability density, peaked
on the corresponding expectation value
\be
\mathcal{P}^>_{\rm H}(\Rh)
\simeq 
\delta\left(\Rh-\braket{\hRh^>}\right)
\ .
\ee
This last estimate follows from the fact that we consider black holes of astrophysical size,
which means that $N$ is a huge number and the uncertainty over the horizon radius 
is negligible, from a collective point of view, contrarily to the single-particle case analysed in \eg~\cite{Casadio:2013tma}.

	The fact that the vast majority of the toy gravitons is sharply distributed, at least from
a macroscopic perspective, has a simple consequence on the expectation value of the \emph{horizon operators} \eqref{O>} and \eqref{O<}.
Since the cut-off $\ell$ is a non-operatorial quantity, 
coming from an effective quantum-gravitational description, with the aid of Eq.~\eqref{ADM}
it is straightforward to infer $\braket{\wh{R}^>_{\rm H}} = \Rh^>$ once the constraint~\eqref{HorConstr>} has been implemented. Obviously, the same holds for $\expec{\hat{R}^<_{\rm H}}$.

	The sought probability reads
\be
P_{\rm BH}(\ell;N)
\,=\,
\prod_{i=1}^N
P_{\rm S} \left( r_i<\expec{\hRh^>} \right)
\,=\,
\left[
P_{\rm S} \left( r<\expec{\hRh^>} \right)
\right]^N
\label{PBHellN}
\ee
with a sufficient degree of accuracy. 
\par
Taking profit of Eq.~\eqref{Rh>} and recalling the definition of the lower-incomplete
Euler Gamma function
\be
\gamma(s,x)
\,=\,
\int_0^x \d t \, t^{s-1} \, e^{-t}
\ ,
\ee
we can easily compute
\be
P _{\rm S} \left( r<\expec{\hRh^>} \right)
&=&
4 \, \pi \, \int_0^{\Rh^>} \d r  \, r^2 \, |\phi_0(r)|^2 
\notag
\\
&=&
\frac{2}{\sqrt{\pi}}\,\gamma\left(\frac{3}{2},\,\frac{4 \, n_0+3}{N} \, \frac{(\Rh^>)^2}{\lp^2}\right)
\notag
\\
&=&
\frac{2}{\sqrt{\pi}}\,
\gamma\left[\frac{3}{2},\frac{4\,n_0+3}{N} \, \left(\frac{2 \, M}{\mpl} -
\frac{\mpl}{2 \, M}\frac{\ell^2}{\lp^2}\right)^2\right]
\ee
and expand the result to the leading order in the cut-off $\ell$
\be
P _{\rm S} \left( r<\expec{\hRh^>} \right)
&=&
\frac{2}{\sqrt{\pi}}\gamma\left(\frac{3}{2},\frac{4(4n_0+3)}{N}\frac{M^2}{\mpl^2}\right)
\,\times \notag
\\
&&\quad
-\frac{8\,M}{\sqrt{\pi}\mpl}\left(\frac{4n_0+3}{N}\right)^{3/2}\, \exp\left(-\frac{4(4n_0+3)M^2}{N\mpl^2}\right)\,\frac{\ell^2}{\lp^2} \notag \\ 
&& + \mathcal{O}\left(\frac{\ell^3}{(\lp M/\mpl)^3}\right)
\ .
\ee
The total probability~\eqref{PBHellN} is therefore given by
\be
P_{\rm BH}(\ell;N)
=
\left\{
\frac{2}{\sqrt{\pi}}\,
\gamma\left[\frac{3}{2},\frac{4\, n_0+3}{N}\left(\frac{2\, M}{\mpl}-
\frac{\mpl}{2 \, M}\frac{\ell^2}{\lp^2}\right)^2\right]
\right\}^N
\ .
\ee
Expanding for $\mpl \, \ell / M \, \lp \ll 1$ (as above) and plugging in the scaling relation $M=\sqrt{N} \, \mpl$, we finally have
\be
P_{\rm BH}(\ell;N)
&\simeq &
\left[
\frac{2}{\sqrt{\pi}}\,
\gamma\left(\frac{3}{2},4(4n_0+3)\right)
\right]^{N-1}\, \times
\notag
\\
&&\quad
\left[
\frac{2}{\sqrt{\pi}}\,
\gamma\left(\frac{3}{2},4(4n_0+3)\right)
-\frac{8(4n_0+3)^{3/2}}{\sqrt{\pi}N}\, e^{-4(4n_0+3)}\,\frac{\ell^2}{\lp^2}
\right]
\label{eq:PBHN}
\ .
\ee
It is interesting to see that the difference
\be
\label{sigma1}
\sigma(\ell)
&=&
|P_{\rm BH}(0;N)-P_{\rm BH}(\ell;N)|
\notag
\\
&\sim &
\frac{8 \, (4\, n_0+3)^{3/2}}{\sqrt{\pi}N}\,
\left[
\frac{2}{\sqrt{\pi}}\,
\gamma\left(\frac{3}{2},4 \,(4 \, n_0+3)\right)
\right]^{N-1}
\exp\left(-4 \, (4 \,n_0+3)\right) \,\frac{\ell^2}{\lp^2} \, ,
\ee
signals the breaking of a Schwarzschild-like black hole configuration, due to QG effects 
at the center, when $\sigma(\bar \ell) \sim 1$,
for a certain value of the cut-off $\bar \ell$, which is nonetheless constrained to respect
the bound Eq.~\eqref{eq:Bound2HorCrit}.

	Recalling that $\gamma (3/2, x)$ is a positive and (strictly) monotonically increasing function on $x>0$ with $\sup _{x>0} \{\gamma (3/2, x)\} = \sqrt{\pi} / 2 < 1$, it is easy to see that 
	\be
	(2/\sqrt{\pi}) \, \gamma \left(\frac{3}{2}, 4 \,(4 \, n_0+3) \right)
	\,<\,
	1
	\ee
	 and positive for all (fixed) $n_0$. Moreover, from the assumption $\mpl \, \ell / M \, \lp \ll 1$ one can easily infer that
	\be 
	\frac{\ell}{\lp} \ll \frac{M}{\mpl} \sim \sqrt{N} \, . 
	\ee
	Thus, plugging these considerations in Eq.~\eqref{sigma1}, we get
	\be 
	\sigma(\ell) \ll \frac{8 \, (4\, n_0+3)^{3/2}}{\sqrt{\pi}}\,
\left[
\frac{2}{\sqrt{\pi}}\,
\gamma\left(\frac{3}{2},4 \,(4 \, n_0+3)\right)
\right]^{N-1}
\exp\left(-4 \, (4 \,n_0+3)\right) \ll 1
	\ee
for all (fixed) value of $n_0$ and $\forall N \gg 1$.	Therefore, there are no signals of
breakdown of the black hole configuration caused by quantum fluctuations, since $\sigma(\ell)\ll 1$.

Concerning the inner horizon, the same argument translates to 
\be
P _{\rm S} \left(r<\expec{\hRh^<}\right)
=
\frac{2}{\sqrt{\pi}}\,
\gamma\left[\frac{3}{2},\frac{4n_0+3}{N}
\left(\frac{\ell}{\lp} + \frac{\mpl\ell^2}{4\lp^2 M} \right)^2\right]
\ ,
\ee
which yields
\be
P_{\rm IH}(\ell;N)
\simeq
\left\{\frac{2}{\sqrt{\pi}}\,
\gamma\left[\frac{3}{2},\frac{4n_0+3}{N}
\left(\frac{\ell}{\lp} + \frac{\ell^2}{4\sqrt{N}\lp^2} \right)^2\right]\right\}^N
\label{eq:PIHN}
\ .
\ee
Again, because of the condition $\mpl \, \ell / M \, \lp \ll 1$, together with the strict monotonicity of $\gamma (3/2, x)$, it is easy to see that
\be
P_{\rm IH}(\ell;N)
\ll
\left\{
\frac{2}{\sqrt{\pi}}\,
\gamma\left[\frac{3}{2} \, , \, \frac{25}{16} \, \left( 4 \, n_0+3 \right) \right]
\right\}^N \ll 1
\ee
for all (fixed) value of $n_0$ and $\forall N \gg 1$. 
This shows that the probability that our static configuration
has an internal horizon is negligible. 

In particular, it has been proposed~\cite{Casadio:2015lis} that $N\to \infty$ is the limit, in which a purely classical geometry shall emerge from the effective description of the corpuscular black hole model. This statement is enforced by the present analysis, since
\be
\lim _{N \to \infty} P_{\rm IH} (\ell ; \, N)
= 0
\ee
means that the center of the inner region is not screened by any horizon, as in the vacuum solutions in GR.
\section{Conclusions and outlook} \label{sec-4}
In this work, we analysed the possible ramifications of applying HQM to non-singular,
static metrics.
First of all, an observer outside the trapped region sees little or no deviation from 
Schwarzschild's geometry, even for what concerns the quantum fluctuations over its characteristic
features, although the interior is drastically different. 
This outcome strongly supports the standard knowledge, according to which 
the central singularity is an artifact of the incompleteness of GR at describing short-scale
gravity and not some strange (and appalling) property of nature.
The only information about the purely quantum nature of gravitation is encoded
in the cut-off $\ell$ alone, which still leaves the freedom to motivate regularity at the origin 
with a wide variety of microscopic theories.
On the other hand, it is of paramount importance to note that the probability to have an inner
horizon Eq.~\eqref{eq:PIHN} is always negligible within the framework of the corpuscular model of black holes. This hints at the fact that the formation of a non-singular black hole of the Hayward-type is rather unlikely in this scheme.
\par
The reader should bear in mind, however, that we just considered an already formed 
self-sustained gravitating system and checked to which extent quantum-gravitational
fluctuations can break this configuration. We have not been able to compute the
probability to take a lump of baryonic matter and get an event horizon,
once it has collapsed in some region of space, \eg as in Ref~\cite{Casadio:2015qaq,Casadio:2013tma}.
Unfortunately, it is indeed impossible to replicate this machinery in any multi-particle model 
without requiring a fully dynamical approach, which extends 
the idea of Ref.~\cite{Casadio:2014twa} including somehow 
the quantum details of gravitational collapse. This colossal assignment is left for future research.
\section*{Acknowledgements}
A.~Giugno and A.~H. were supported by the ERC Advanced Grant 339169 ``Selfcompletion",
during the realisation of this paper. A.~Giusti is partially supported by the INFN grant FLAG.
The research activity of A.~Giugno and A.~Giusti has also been carried out in the framework
of activities of the National Group of Mathematical Physics (GNFM, INdAM).


\begin{thebibliography}{99}
\bibitem{ft1}
For the sake of completeness: the singularity theorems tell us nothing about the nature of this singularity (for a recent review on the topic, see~\cite{Singularity}), 
indeed they are not forced to be curvature singularities if we move away 
from the spherically symmetric case.
\bibitem{ft2}
We write $\hbar=\lp\mpl$ and $G_{\rm N}=\lp/\mpl$ in Planck units. Nonetheless, we assume $c=1$.
\end{thebibliography}
\end{document}